\documentclass[paper,twocolumn,showpacs,superscriptaddress]{revtex4-1}
\usepackage{amsfonts}
\usepackage{amsmath}
\usepackage{amssymb}
\usepackage{graphicx}
\begin{document}

\title{Perfect light trapping in nanoscale thickness semiconductor films with resonant
back reflector and spectrum-splitting structures}
\author{Jiang-Tao Liu}
\email{jtliu@semi.ac.cn}
\address{Nanoscale Science and Technology  Laboratory,  Institute for Advanced Study, Nanchang University, Nanchang 330031, China}
\address{Department of Physics, Nanchang University, Nanchang 330031, China}

\author{Xin-Hua Deng}
\address{Department of Physics, Nanchang University, Nanchang
330031, China}

\author{Wen Yang}
\address{Beijing Computational Science Research Center, Beijing 100084, China}

\author{Jun Li}
\address{Department of Physics, Semiconductor Photonics Research Center, Xiamen University, Xiamen 361005,
China}
\date{\today }

\begin{abstract}
The optical absorption of nanoscale thickness semiconductor films on top of
light-trapping structures  based on optical interference effects combined with spectrum-splitting structures is
theoretically investigated. Nearly perfect absorption over a broad spectrum
range can be achieved in $<100$ nm thick films on top of one-dimensional
photonic crystal or metal films. This phenomenon can be attributed to
interference induced photonic localization, which enhances the absorption
and reduces the reflection of the films. Perfect solar absorption and low
carrier thermalization loss can be achieved when the light-trapping
structures  with wedge-shaped spacer layer or semiconductor films are combined with spectrum-splitting structures.
\end{abstract}

\maketitle

\address{Nanoscale Science and Technology  Laboratory,  Institute for
Advanced Study, Nanchang University, Nanchang 330031, China} %
\address{Department of Physics, Nanchang University, Nanchang 330031, China}

\address{Department of Physics, Nanchang University, Nanchang
330031, China}

\address{Beijing Computational Science Research Center, Beijing 100084,
China}

\address{Department of Physics, Semiconductor Photonics Research Center, Xiamen University, Xiamen 361005,
China}

To reduce carbon dioxide release and cope with the increasing energy demand,
photovoltaic solar cells have attracted significant attention. However, the
cost efficiency and power conversion efficiency of solar cells is still
relatively low. The high cost is
mainly due to the use of expensive semiconductor materials, such as thick
crystalline silicon films \cite{B98AG,B13IMD}. The efficiency is primarily
limited by thermalization loss \cite{B98AG,B13IMD,NM12AP}.

A best way to enhance cost efficiency is to reduce solar cells to nanoscale
\cite{B13IMD,JAP12SM,S13JW,NC13GM,JAP13DBT,NL14SS,NP13PK} to reduce material
usage. In solar cells, the carrier collection length should be shorter than
the carrier diffusion length (i.e., the distance travelled by a carrier
before recombination). Thus nanoscale solar cells can use low-cost materials
with lower carrier diffusion length (e.g., CuO, FeS2, and organic materials)
to further reduces the cost \cite{JAP12SM}. However, nanoscale materials are
usually too thin to completely absorb the solar light. Plasmon, photonic
crystal, and quasi-random nanostructures can enhance absorption \cite%
{JAP12SM,NM12KV,NC13ERM,EES12VR,EES11EEC,APL12AB,OE12MYK,JAP13MR,AM13QG,JMCA14SE,OE12SKK,EES14KXW}. Many theoretical and experimental works on interference-based
light-trapping structures have demonstrated further improvement of the
absorption of ultra-thin films compared with traditional light-trapping
structures \cite{NM13MAK,AM14HS,APL13XLZ} at the cost of reduced working
frequency range. Perfect broadband absorption is still difficult to achieve
in these structures.

Carrier thermalization loss results from energy mismatch between the photon
energy $E_{\omega }=\hbar \omega $ and the bandgap of semiconductors $E_{g}$
\cite{B98AG,B13IMD,NM12AP}. Photons with energy below the bandgap of
semiconductors are not absorbed, while photons with energy above the bandgap
can create only a maximum open-circuit voltage $V_{oc}$. The excess energy $%
E_{\omega }-eV_{oc}$ is converted to heat loss, where $e$ is the electron
charge \cite{B98AG,B13IMD,NM12AP}. Multijunction solar cells can reduce
carrier thermalization loss with increased number of junctions. However,
only 2- and 3-junction solar cells can be used for industrial production due
to the limitation of lattice matching and current matching requirement \cite%
{B13IMD,NM12AP}. Another widely studied and demonstrated technque to reduce
the thermalization loss is to use side-by-side subcells and a
spectrum-splitting structure \cite{NM12AP,SEM04AGI,PP09AB,PP10MAG}, which
directs solar light of different wavelengths to different subcells with
different bandgaps $E_{g}$. The thermalization loss can be reduced to
approximately 10\% by using 8-10 different subcells \cite{NM12AP}.

\begin{figure}[b]
\centering
\includegraphics[width=0.95\columnwidth,clip]{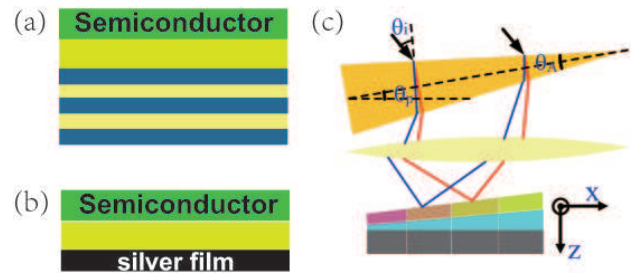}
\caption{(Color online) Schematic of SFs on (a) 1DPC and (b) Ag films with a
spacer layer. (c) Schematic of spectrum-splitting structure combined with
light-trapping structure. }
\label{fig1}
\end{figure}

In this letter, we show that by combining two distinct well-established
techniques, the interference-based light-trapping structures and the
spectrum-splitting structure, perfect broadband solar absorption can be
achieved in nanoscale thickness semiconductor films (SFs). First, for the 32nm GaAs
films on top of a light-trapping structure, numerical and analytical results
show narrow-band resonant absorption exceeding 90\% ($\sim $ nine times
larger than that of suspended GaAs films) as a result of photon
localization. Second, we match the resonant absorption of the light-trapping
structure with the spectrum-splitting structure by using wedge-shaped spacer layer or SFs and demonstrates perfect
broadband solar absorption in $30-240$ nm thick SFs.

The light-trapping structure consists of a SF on top of a spacer layer and a
one-dimensional photonic crystal (1DPC) [Fig. 1(a)] or 130 nm thick Ag film
[Fig. 1(b)] at the bottom as the resonant back reflector. As demonstrated
previously, such structure without wedge-shaped spacer layer or SFs can greatly
enhance the absorption of graphene,
MoS$_{2}$, and ultra-thin SFs within a certain wavelength region\cite%
{AM14HS,APL13XLZ,APL07KC,APL12JTL,EPL13NMRP,NL12BSR,APL12DH,JAP14JTL}. We also should point out the physical mechanism  is different with the proposed  structures consisting of thick SFs on photonic crystal \cite{OE07PB,OE08JGM}. The optical interference effects in thick SFs  are unimportant due to the strong absorption and an antireflection coating layer must be used to reduce the reflection.   The
1DPC is composed of 8.5 periods of alternating doped ZnS and SiO$_{2}$
layers. The refractive indices of ZnS and SiO$_{2}$ at $\lambda =550$ nm are
$n_{ZnS}=2.59$ and $n_{SiO_{2}}=1.55$, respectively. The permittivities for
silver film and SFs are frequency dependent \cite{AP85EDP,B99SA}. All layers
are nonmagnetic ($\mu =1$). The thickness of the ZnS (SiO$_{2}$) layers is $%
\lambda _{pcs}/4n_{ZnS}$ ($\lambda _{pcs}/4n_{SiO_{2}}$), where $\lambda
_{pcs}$ is the center wavelength. The spectrum-splitting structure is
composed of a dense flint ZF13 prism and a low-dispersion convex lens [Fig.
1(c)]. The refractive indices for ZF13 glass is $n_{ZF13}^{2}=z_{g1}-z_{g2}%
\lambda ^{2}+z_{g3}\lambda ^{-2}+z_{g4}\lambda ^{-4}-z_{g5}\lambda
^{-6}+z_{g6}\lambda ^{-8}$, where $z_{g1}=3.05344,$ $z_{g2}=1.2752\times
10^{-2},z_{g3}=4.0609\times 10^{-2},z_{g4}=2.2706\times
10^{-3},z_{g5}=7.8087\times 10^{-5}$, and $z_{g6}=1.9874\times 10^{-5}$.

To model the absorption of SFs in this structure, the standard transfer
matrix method is used \cite{APL12JTL,JAP14JTL}.
In the \emph{l}th layer, the electric field of the TE mode light with incident angle $\theta_{i}$ is given by
\begin{equation}\mathbf{E}_{l}(z,y)=\left[  A_{l}e^{ik_{lz}%
\left(  z-z_{l}\right)  }+B_{l}e^{-ik_{lz}\left(  z-z_{l}\right)  }\right]
e^{ik_{ly}y}\mathbf{e}_{x}, \label{TMM:a1}\end{equation}
and the magnetic field of the TM mode in the \emph{l}th layer is given by
\begin{equation}\mathbf{H}_{l}(z,y)=\left[  A_{l}e^{ik_{lz}%
\left(  z-z_{l}\right)  }+B_{l}e^{-ik_{lz}\left(  z-z_{l}\right)  }\right]
e^{ik_{ly}y}\mathbf{e}_{x},\end{equation}
where $k_{l}=k_{lr}+ik_{li}$ is the wave vector of the light, $\mathbf{e}_{x}$ is the unit vectors in the x direction, and $z_{l}$ is the position of the \emph{l}th layer in the z direction.

The electric fields of TE mode or the magnetic fields of TM mode in the (\emph{l}+1)th layer are related to the incident fields by the transfer matrix utilizing the boundary condition
\begin{equation}\binom{A_{l+1}}{B_{l+1}}=\left(
\begin{array}
[c]{cc}%
T_{11} & T_{12}\\
T_{21} & T_{22}%
\end{array}
\right)  \binom{A_{0}}{B_{0}}.\end{equation}

Thus, we can obtain  the absorbance of \emph{l}th layer $\mathcal{A}_{l}$  using the Poynting vector $\textbf{S}=\textbf{E}\times\textbf{H}$ \cite{APL12JTL,JAP14JTL}
\begin{equation}
\mathcal{A}_{l}=[\mathcal{S}_{(l-1)i}+\mathcal{S}_{(l+1)i}-\mathcal{S}_{(l-1)o}-\mathcal{S}_{(l+1)o}]/\mathcal{S}_{0i},
\end{equation}
where  $\mathcal{S}_{(l-1)i}$ and $\mathcal{S}_{(l-1)o}$ [$\mathcal{S}_{(l+1)i}$ and $\mathcal{S}_{(l+1)o}$] are the incident  and  outgoing Poynting vectors (\emph{l}-1)th [(\emph{l}+1)th] layer, respectively, $\mathcal{S}_{0i}$ is the incident Poynting vectors in air.  For a perfect reflecting
mirror without a spacer layer, straightforward algebra gives the absorptance
\begin{equation}
\mathcal{A}_{po}=1-|(1+\gamma _{sm})/(1-\gamma _{sm})|^{2},  \label{ab_pf}
\end{equation}%
where $n_{sm}\gamma _{sm}=(1-\mathfrak{p}^{2})/{(1+\mathfrak{p}^{2})}$; $%
\mathfrak{p}=e^{ik_{sm}d_{sm}}$; $n_{sm}$ and $d_{sm}$ are the refractive
index and thickness of SFs, respectively; and $k_{sm}$ is the wave vector in
the SFs. For thick metal film mirrors without the spacer layer, the
absorptance
\begin{equation}
\mathcal{A}_{mf}=1-|1-\xi_{sm}-\zeta_{sm}|^{2}-n_{sm}|\xi_{sm}\mathfrak{p}+\zeta_{sm}%
\mathfrak{p}^{\ast }|^{2},  \label{ab_mf}
\end{equation}%
where $\xi_{sm}=2n_{21}\mathfrak{p}^{2}/(n_{21}n_{10}\mathfrak{p}%
^{2}-n_{01}n_{12})$, $\zeta_{sm}=2n_{12}/(n_{12}n_{01}-n_{10}n_{21}\mathfrak{p}%
^{2})$, $n_{21}=n_{mf}-n_{sm}$, $n_{12}=n_{mf}+n_{sm}$, $n_{10}=1-n_{sm}$,
and $n_{01}=n_{sm}+1$, $n_{mf}$ are the refractive indices of metal films.

\begin{figure}[!t]
\centering
\includegraphics[width=0.95\columnwidth,clip]{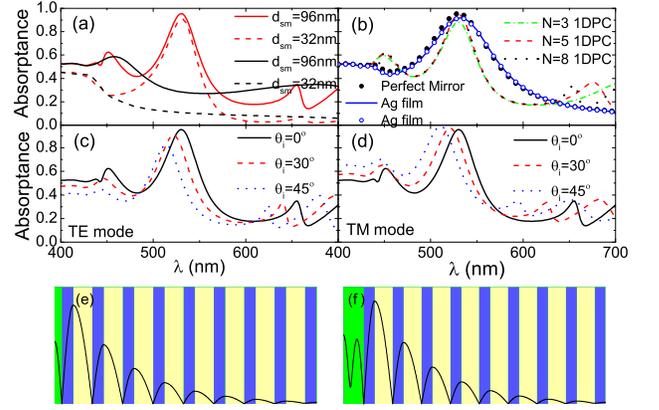}
\caption{(Color online) (a) Absorptance of different thickness GaAs films
without a spacer layer as a function of wavelength for suspended GaAs films
(black line) and GaAs films with a 1DPC (red or grey line). (b) Absorptance
of 96 nm GaAs films on 1DPC as a function of wavelength with different
period numbers N of 1DPC. The black solid-circle curve shows the absorptance
of 96 nm GaAs films with the perfect reflection mirror. The blue solid line
and blue open circle curve shows the absorptance of 78 nm GaAs films on Ag
films. The absorptance of 96 nm GaAs films on a 1DPC as a function of light
wavelength for different incident angles for (c) TE mode and (d) TM mode.
Optical field distribution in (e) 32 and (f) 96 nm GaAs films on 1DPC.}
\label{fig2}
\end{figure}

For the spectrum-splitting structure, the angle of refraction of light through a prism is given by
\begin{equation}
	\sin[\theta_{rp}(\lambda)]=\sin\{\theta_{A}-\arcsin[\sin(\theta_{ip})/n_{ZF13}(\lambda)]\}n_{ZF13}(\lambda),
\end{equation}
where $\theta_{A}$ is the prism vertex angle, $\theta _{ip}$ is the light
incidence angle to the prism. The the light
incidence angle to the convex lens is $\theta _{il}(\lambda)=\theta_{rp}(\lambda)-\theta_{A}/2-\theta _{p}$,
where $\theta _{p}$ is the prism tilt angle. For perfect lens, the light are focused at $x(\lambda)=f_{0}\tan[\theta _{il}(\lambda)]$, where $f_{0}$ is the focal length of the lens.
For actual convex lens with aberrations, each of the refraction of light through the lens can be
solved numerically using the Snell's Law.

As shown in Fig. 2(a), for 32 nm (96 nm) GaAs films, the maximum absorbance
at the resonant wavelength $\lambda =530$ nm of solar radiation is improved
by the 1DPC from $\sim $10.8\% (28.4\%) for suspended films to $\sim $90.8\%
(95.5\%) even without antireflection coating layers. The magnitude of the enhancement is much larger than that of thick SFs on photonic crystal even with  antireflection coating layers\cite{OE07PB,OE08JGM}. This is because the SFs act as a 1D surface defect that leads to
localization on the surface due to interference effects [Figs. 2(e) and 2(f)], thereby reducing the
reflection and enhancing the absorption \cite{APL12JTL,NL12BSR,JAP14JTL},
similar to graphene and monolayer MoS$_{2}$ on top of 1DPC. As shown in Fig.
2(b), the 1DPC with increasing number $N$ of periods has increasing
reflectivity, photonic localization, and hence absorption. The SF resonant
absorption with the 1DPC is close to that with a perfect reflection mirror
(black solid circles) and slightly better than that with a Ag film (blue
solid line and blue open circle), which has a lower reflectivity. However,
the width of the resonant absorption for 1DPC is smaller than that for Ag
film, due to the limited photonic band gap of the 1DPC. Away from the
resonant absorption, the SF absorption on the Ag film almost coincides with
that on a perfect reflection mirror.

For oblique light incidence, the large refractive index of GaAs ensures, by
Snell's law, that the light propagation angle $\theta ^{\prime }$ in the
GaAs film is small even with a large incident angle $\theta _{i}$. Thus the
resonant absorption wavelength $\propto \cos \theta ^{\prime }$ for the SFs
on the 1DPC is less affected by the light incident angle [Figs. 2(c) and
2(d)] and favors the design of solar cells.

An essential ingredient of our proposal is the tunability of the resonant
absorption wavelength $\lambda _{\mathrm{R}}$ of the light-trapping
structure. As shown in Figs. 3(a) and 3(b), for the 1DPC, $\lambda _{\mathrm{%
R}}=4n_{sm}d_{sm}/(2m+1)$ ($m\in \mathbb{Z}$) decreases with decreasing SF
thickness $d_{sm}$. This remains qualitatively true for the Ag film, since
we still have the estimate $\lambda _{\mathrm{R}}\sim 4n_{sm}d_{sm}/(2m+1)$.
Another approach to tune $\lambda _{\mathrm{R}}$ is to insert a transparent
(e.g., AlAs) spacer layer, e.g., for the 100 nm GaAs film on the 1DPC with $%
\lambda _{pcs}=700$ and Ag films, $\lambda _{\mathrm{R}}$ almost linearly
increases with the thickness of the spacer layer [Figs. 3(c) and 3(d)]. The
spacer layer can also act as the buffer layer to enhance or reduce the
lattice distortion of the SFs.

\begin{figure}[!t]
\centering
\includegraphics[width=0.95\columnwidth,clip]{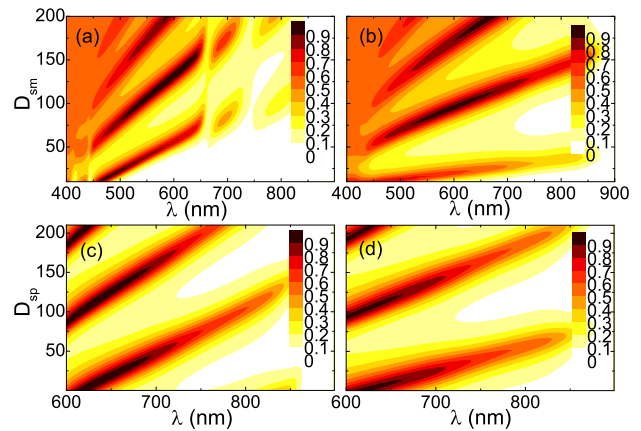}
\caption{(Color online) Contour plots of the absorptance of the GaAs films
on (a) 1DPC and (b) Ag films without spacer layers as a function of light
wavelength and thickness of GaAs films with $\protect\lambda_{pcs}=550$ nm.
Contour plots of the absorptance of 100 nm GaAs films on (c) 1DPC and (d) Ag
films as a function of light wavelength and thickness of AlAs spacer layers.}
\label{fig3}
\end{figure}

The problem with a given light-trapping structure is that enhanced
absorption appears only within a certain range of wavelength. To cure this
problem, we propose to combine it with the spectrum-splitting structure,
which has been widely used experimentally to reduce the carrier
thermalization loss in solar cells \cite{NM12AP,SEM04AGI,PP09AB,PP10MAG}.
With the spectrum-splitting structure, a dense flint ZF13 prism and a
low-dispersion convex lens focuses solar light with different wavelengths
onto different locations on the SFs of the light-trapping structure [Fig.
1(c)]. By appropriately choosing the species and thicknesses of SFs or the
thicknesses of the spacer layers to match the solar wavelength on each
location (i.e., use the wedge-shaped spacer layer or SFs), perfect absorption can be achieved for each wavelength. Since the
slope of the SFs is smaller than $2.5\times 10^{-5}$, the SF surface can be
treated as parallel planes at each point. Thus, the transfer matrix method
can be used in the calculation. The parameters include [Fig.
1(c)]: the prism vertex angle $\theta _{A}=30^{\circ }$, the light
incidence angle $\theta _{i}=26.4^{\circ }$ (the refracted light in the
prism is parallel to bottom plane of the prism), the prism tilt angle $%
\theta _{p}=10^{\circ }$ (the light is nearly parallel to z direction), and the refractive indices and spherical radius of
the convex lens are 1.5 and 0.6 m, respectively. CdS ($E_{g}=2.42$ eV), Ga$%
_{0.5}$In$_{0.5}$P($E_{g}=1.9$ eV), GaAs($E_{g}=1.42$ eV), and In$_{0.53}$Ga$%
_{0.47}$As ($E_{g}=0.86$ eV) films consist of side-by-side SFs.

\begin{figure}[t]
\centering
\includegraphics[width=0.91\columnwidth,clip]{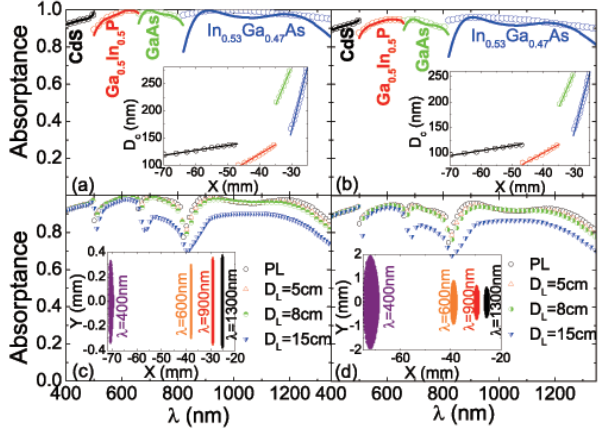}
\caption{(Color online) Absorptance of side-by-side SFs on (a) 1DPC and (b)
Ag films for the SFs with thickness that perfectly match the
spectrum-splitting structure (open circle curves) and for SFs with linearly
varying thicknesses (solid lines). The inset shows the thickness of
side-by-side SFs as a function of the x coordinate. The absorptance of
side-by-side SFs on (c) 1DPC and (d) Ag films with perfect lens (black open
circle curves) and convex lens with different diameters $D_{L}$. The inset
in (c) and (d) shows a focused light spot for different wavelengths for $%
D_{L}=8$ cm and $D_{L}=15$, respectively. }
\label{fig4}
\end{figure}

Numerical results are shown in Fig. 4. When the thickness of SFs perfectly
matches the spectrum splitting with various $\lambda_{pcs}$ (i.e., for
lights with wavelength $\lambda_{0}$ focused at $x=x_{0}$ point, the SF
thickness at $x=x_{0}$ is chosen to obtain maximum absorption), the
absorption of side-by-side SFs is almost always $>90$\% and the maximum
absorption can be $>99$\%. Furthermore, more types of SFs can be used in
this structure because the lattice matching and current matching
requirements are not restricted. Thus, the proposed structures can
significantly reduce carrier thermalization loss. Apart from applications in
solar cells, perfect wide-range absorption can also have applications in
optoelectronic devices, such as photoelectric detectors. Different for
nanowire array solar cells, the proposed structures have low surface
recombination loss. For SFs with linearly varied thickness [solid line in
the inset of Figs. 4(a) and 4(b)], only the absorption of In$_{0.53}$Ga$%
_{0.47}$As film significantly decreases because the dispersion of ZF13 glass
is too small to effectively separate different-wavelength lights within this
wavelength range.

The influence of lens aberrations on SFs absorption is shown in Figs. 4(c)
and 4(d). An actual convex lens cannot focus parallel light to a perfect
point because of lens aberrations, which increase with increased lens
diameter $D_{L}$ [inset of Figs. 4(c) and 4(d)]. Lens aberrations reduce the
light-trapping effect because obtaining the maximum absorption from the
entire whole focused light spot is difficult by varying the thickness of
SFs. The influence of lens aberrations is reduced when spectrum-splitting
structure with strong dispersion ability is used. The changes in SFs
absorption are very small for $D_{L}<8$ cm. The absorption of GaAs and In$%
_{0.53}$Ga$_{0.47}$As significantly decreases for $D_{L}=15$ cm because of
the low dispersion of ZF13 glass within the long wavelength range.

Perfect absorption can be achieved in the full solar spectrum by varying the
spacer-layer thickness. Side-by-side SFs consist of 80 nm CdS, 80 nm Ga$%
_{0.5}$In$_{0.5}$P, 240 nm GaAs, and 240 nm In$_{0.53}$Ga$_{0.47}$As films.
The refractive indices of a spacer layer is 2.6. The absorptance of SFs
combined with spectrum-splitting structure can exceed 90\% by varying
spacer-layer thickness [Fig. 5(a)]. The thickness of SFs do not have to
satisfy $d_{sm}=(2m+1)\lambda/4n_{sm}$. SFs with a large absorption
coefficient can be thinner. The influence of indium tin oxide transparency electrode on
top of SFs is shown in Fig. 5(b). The absorptance of 20 nm indium tin oxide films is
about 5\%, and the absorptance of SFs shows about 5\% reduction.

\begin{figure}[t]
\centering
\includegraphics[width=0.85\columnwidth,clip]{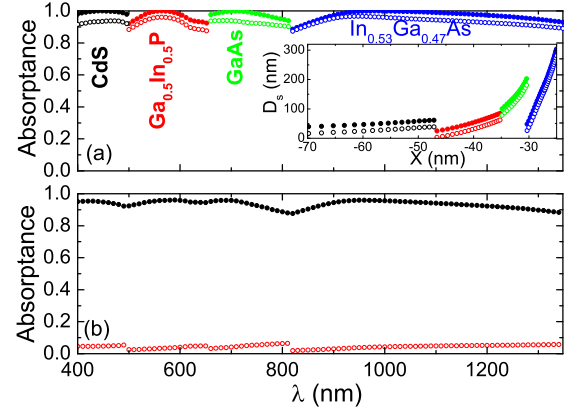}
\caption{(Color online)(a) Absorptance of SFs on 1DPC (solid circle curves)
and Ag films (open circle curves) combined with spectrum-splitting structure
with various spacer layer thickness. The inset shows the spacer-layer
thickness as a function of the x coordinate. (b) Absorptance of SFs (black
solid circle curves) and indium tin oxide  films (red open circle curves) on 1DPC. }
\label{fig5}
\end{figure}

Finally, we discuss on the fabrication techniques and the potential
improvement of the proposed structures. The length of the subcells is about
0.5-2 cm. The subcells can be fabricated separately and then joined using
fasteners, adhesive, or welding. In each subcells, for SFs
on top of 1DPC (which can be regarded as a half of the semiconductor
microcavity) or metal films with spacer layers, similar structure have been
fabricated in experiments \cite{AM14HS}. The difference in this letter is
that the SFs or spacer layers are wedge-shaped layers, which can be grown by
existing technology such as molecular beam epitaxy. As
shown by J. P. Prineas et al.\cite{APL06JPP}, to reduce the speed of light
210 wedge-shaped semiconductor layers are grown by molecular beam epitaxy.
Even though strong light-trapping effects is observed in the proposed
structures, to achieve perfect absorption in $<20$ nm SFs, light-trapping
structure with more strong photonic localization such as the microcavity
should be used \cite{PRB12AF,OL12MAV}. For the proposed spectrum-splitting
structure, dispersion ability within the long wavelength range is low
because of the low dispersion of ZF13 glass that degrades performance. Prism
and convex lens without an antireflection film can reflect about 10-15\%
solar light. Thus, various types of spectrum-splitting structure should be
studied in the future \cite{NM12AP,SEM04AGI,PP09AB,PP10MAG}.

In conclusion, the optical absorption of SFs on top of 1DPC and Ag films
combination with spectrum-splitting structure is investigated. In these
structures, the maximum optical absorptance of $<$100 nm SFs can be $>90\%$
because of photon localization. The absorption of SFs on 1DPC is less
affected by the incident angle of light and can be tuned by varying
thickness of the SFs and spacer layers. By using side-by-side nanoscale thickness  SFs
and combination with spectrum-splitting structure, perfect solar absorption
can be achieved with low carrier thermalization loss. Our proposal can have
significant applications in the development of ultra-thin and
high-efficiency solar cells and in optoelectronic devices, such as
photoelectric detector.

We would like to thank Nianhua Liu for fruitful discussion. This work was
supported by the NSFC (Grant Nos. 11364033, 11274036, 11322542, and 11104232), the MOST (Grant No. 2014CB848700), 
the NSF from the Jiangxi Province (Grant No. 20122BAB212003), 
and Science and Technology Project of Education Department of Jiangxi Province (Grant No. GJJ13005).


\begin{thebibliography}{39}

\bibitem{B98AG}
Adolf Goetzberger and Joachim Knobloch, editors.
\newblock {\em Crystalline Silicon Solar Cells}.
\newblock John Wiley \& Sons, Chichester, 1998.

\bibitem{B13IMD}
I.~M. Dharmadasa, editor.
\newblock {\em Advances in thin-film solar cells}.
\newblock Pan Stanford Publishing, Boca Raton, 2012.

\bibitem{NM12AP}
Albert Polman and Harry~A. Atwater.
\newblock Photonic design principles for ultrahigh-efficiency photovoltaics.
\newblock {\em Nature Materials}, 11:174--177, 2012.

\bibitem{JAP12SM}
S.~Mokkapati and K.~R. Catchpole.
\newblock Nanophotonic light trapping in solar cells.
\newblock {\em J. Appl. Phys.}, 112:101101, 2012.

\bibitem{S13JW}
Jesper Wallentin, Nicklas Anttu, Damir Asoli, Maria Huffman, Ingvar {\AA}erg,
  Martin~H. Magnusson, Gerald Siefer, Peter Fuss-Kailuweit, Frank Dimroth,
  Bernd Witzigmann, H.~Q. Xu, Lars Samuelson, Knut Deppert, and Magnus~T.
  Borgstr\"{o}.
\newblock Plasmonic-enhanced organic photovoltaics: Breaking the 10\%
  efficiency barrier.
\newblock {\em Science}, 339:1057--1060, 2013.

\bibitem{NC13GM}
Giacomo Mariani1, Adam~C. Scofield, Chung-Hong Hung, and Diana~L. Huffaker.
\newblock Gaas nanopillar-array solar cells employing in situ surface
  passivation.
\newblock {\em Nature Communications}, 4:1497, 2013.

\bibitem{JAP13DBT}
Daniel~B. Turner-Evans, Christopher~T. Chen, Hal Emmer, William~E. McMahon, and
  Harry~A. Atwater.
\newblock Optoelectronic analysis of multijunction wire array solar cells.
\newblock {\em J. Appl. Phys.}, 114:014501, 2013.

\bibitem{NL14SS}
Sunil Sandhu, Zongfu Yu, and Shanhui Fan.
\newblock Detailed balance analysis and enhancement of open-circuit voltage in
  single-nanowire solar cells.
\newblock {\em Nano Lett.}, 14:1011--1015, 2014.

\bibitem{NP13PK}
Peter Krogstrup, Henrik~Ingerslev J{\o}gensen, Martin Heiss, Olivier Demichel,
  Jeppe~V. Holm, Martin Aagesen, Jesper Nygard1, and Anna~Fontcuberta i~Morral.
\newblock Single-nanowire solar cells beyond the shockley¨cqueisser limit.
\newblock {\em Nature Photonics}, 7:306--310, 2013.

\bibitem{NM12KV}
Kevin Vynck, Matteo Burresi, Francesco Riboli, and Diederik S.Wiersma.
\newblock Photon management in two-dimensional disordered media.
\newblock {\em Nature Materials}, 11:1017--1022, 2012.

\bibitem{NC13ERM}
Emiliano~R. Martins, Juntao Li, YiKun Liu, Vale¡ärie Depauw, Zhanxu Chen,
  Jianying Zhou, and Thomas~F. Krauss.
\newblock Deterministic quasi-random nanostructures for photon control.
\newblock {\em Nature Communications}, 4:2665, 2013.

\bibitem{EES12VR}
Veronika Rinnerbauer, Sidy Ndao, Yi~Xiang Yeng, Walker~R. Chan, Jay~J.
  Senkevich, John~D. Joannopoulos, Marin Solja\v{c}i\'{c}ab, and Ivan
  Celanovicb~Show Affiliations.
\newblock Recent developments in high-temperature photonic crystals for energy
  conversion.
\newblock {\em Energy Environ. Sci.}, 5:8815--8823, 2012.

\bibitem{EES11EEC}
Mauricio~E. Calvo, Silvia Colodrero, Nuria Hidalgo, Gabriel Lozano, Carmen
  L\'{o}pez-L\'{o}pez, Olalla S\'{a}nchez-Sobradoa, and Hern\'{a}nM\'{\i}guez.
\newblock Porous one dimensional photonic crystals: novel multifunctional
  materials for environmental and energy applications.
\newblock {\em Energy Environ. Sci.}, 4:4800--4812, 2011.

\bibitem{APL12AB}
A.~Basch, F.~J. Beck, T.~S\"{o}derstr\"{o}m, S.~Varlamov, and K.~R. Catchpole.
\newblock Gaas nanopillar-array solar cells employing in situ surface
  passivation.
\newblock {\em Appl. Phys. Lett.}, 100:243903, 2012.

\bibitem{OE12MYK}
M.~Y. Kuo, J.~Y. Hsing, T.~T. Chiu, C.~N. Li, W.~T. Kuo, T.~S. Lay, and M.~H.
  Shih.
\newblock Quantum efficiency enhancement in selectively transparent silicon
  thin film solar cells by distributed bragg reflectors.
\newblock {\em Opt. Express}, 20:A828--A835, 2012.

\bibitem{JAP13MR}
Ma'ayan Rumbak, Iris Visoly-Fisher, and Rafi Shikler.
\newblock Broadband absorption enhancement via light trapping in periodically
  patterned polymeric solar cells.
\newblock {\em J. Appl. Phys.}, 114:013102, 2013.

\bibitem{AM13QG}
Qiaoqiang Gan, Filbert~J. Bartoli, and Zakya~H. Kafafi.
\newblock Plasmonic-enhanced organic photovoltaics: Breaking the 10\%
  efficiency barrier.
\newblock {\em Adv. Mater.}, 25:2385--2396, 2013.

\bibitem{JMCA14SE}
Sergey Eyderman, Alexei Deinegaa, and Sajeev Johnab.
\newblock Near perfect solar absorption in ultra-thin-film gaas photonic
  crystals.
\newblock {\em Journal of Materials Chemistry A}, 2:761--769, 2014.

\bibitem{OE12SKK}
Sun-Kyung Kim, Kyung-Deok Song, and Hong-Gyu Park.
\newblock Design of input couplers for efficient silicon thin film solar
  absorbers.
\newblock {\em Opt. Express}, 20:A997--A1004, 2012.

\bibitem{EES14KXW}
Ken~Xingze Wang, Zongfu Yu, Victor Liu, Aaswath Raman, Yi~Cui, and Shanhui Fan.
\newblock Light trapping in photonic crystals.
\newblock {\em Energy Environ. Sci.}, 7:2725, 2014.

\bibitem{NM13MAK}
Mikhail~A. Kats, Romain Blanchard, Patrice Genevet, and Federico Capasso.
\newblock Nanometre optical coatings based on strong interference effects in
  highly absorbing media.
\newblock {\em Nature Materials}, 12:20--24, 2013.

\bibitem{AM14HS}
Haomin Song, Luqing Guo, Zhejun Liu, Kai Liu, Xie Zeng, Dengxin Ji, Nan Zhang,
  Haifeng Hu, Suhua Jiang, and Qiaoqiang Gan.
\newblock Nanocavity enhancement for ultra-thin film optical absorber.
\newblock {\em Adv. Mater.}, 26:2737--2743, 2014.

\bibitem{APL13XLZ}
Xu-Lin Zhang, Jun-Feng Song, Xian-Bin Li, Jing Feng, and Hong-Bo Sun.
\newblock Anti-reflection resonance in distributed bragg reflectors-based
  ultrathin highly absorbing dielectric and its application in solar cells.
\newblock {\em Appl. Phys. Lett.}, 102:103901, 2013.

\bibitem{SEM04AGI}
A.G. Imenes and D.R. Mills.
\newblock Spectral beam splitting technology for increased conversion
  efficiency in solar concentrating systems: a review.
\newblock {\em Solar Energy Materials \& Solar Cells}, 84:19--69, 2004.

\bibitem{PP09AB}
Allen Barnett, Douglas Kirkpatrick, Christiana Honsberg, Duncan Moore, Mark
  Wanlass, Keith Emery, Richard Schwartz, Dave Carlson, Stuart Bowden, Dan
  Aiken, Allen Gray, Sarah Kurtz, Larry Kazmerski, Myles Steiner, Jeffery Gray,
  Tom Davenport, Roger Buelow, Laszlo Takacs, Narkis Shatz, John Bortz, Omkar
  Jani, Keith Goossen, Fouad Kiamilev, Alan Doolittle, Ian Ferguson, Blair
  Unger, Greg Schmidt, Eric Christensen, and David Salzman.
\newblock Very high efficiency solar cell modules.
\newblock {\em Prog. Photovolt: Res. Appl.}, 17:75--83, 2009.

\bibitem{PP10MAG}
Martin~A. Green and Anita Ho-Baillie.
\newblock Forty three per cent composite split-spectrum concentrator solar cell
  efficiency.
\newblock {\em Prog. Photovolt: Res. Appl.}, 18:42--47, 2010.

\bibitem{APL07KC}
Kai Chang, J.~T. Liu, J.~B. Xia, and N.~Dai.
\newblock Enhanced visibility of graphene: Effect of one-dimensional photonic
  crystal.
\newblock {\em Appl. Phys. Lett.}, 91:181906, 2007.

\bibitem{APL12JTL}
J.~T. Liu, N.~H. Liu, J.~Li, X.~J. Li, and J.~H. Huang.
\newblock Enhanced absorption of graphene with one-dimensional photonic
  crystal.
\newblock {\em Appl. Phys. Lett}, 101:052104, 2012.

\bibitem{EPL13NMRP}
N.~M.~R. Peres and Yu.~V. Bludov.
\newblock Enhancing the absorption of graphene in the terahertz range.
\newblock {\em EPL}, 101:58002, 2013.

\bibitem{NL12BSR}
Berardi Sensale-Rodriguez, Rusen Yan, Subrina Rafique, Mingda Zhu, Wei Li,
  Xuelei Liang, David Gundlach, Vladimir Protasenko, Michelle~M. Kelly, Debdeep
  Jena, Lei Liu, , and Huili~Grace Xing.
\newblock Extraordinary control of terahertz beam reflectance in graphene
  electro-absorption modulators.
\newblock {\em Nano Lett.}, 12:4518--4522, 2012.

\bibitem{APL12DH}
Haixia Da and Cheng-Wei Qiu.
\newblock Graphene-based photonic crystal to steer giant faraday rotation.
\newblock {\em Appl. Phys. Lett.}, 100:241106, 2012.

\bibitem{JAP14JTL}
Jiang-Tao Liu, Tong-Biao Wang, Xiao-Jing Li, and Nian-Hua Liu.
\newblock Enhanced absorption of monolayer mos2 with resonant back reflector.
\newblock {\em J. Appl. Phys.}, 115:193511, 2014.

\bibitem{OE07PB}
Peter Bermel, Chiyan Luo, Lirong Zeng, Lionel~C. Kimerling, and John~D.
  Joannopoulos.
\newblock Improving thin-film crystalline silicon solar cell efficiencies with
  photonic crystals.
\newblock {\em Opt. Express}, 15:16986, 2007.

\bibitem{OE08JGM}
James~G. Mutitu, Shouyuan Shi, Caihua Chen, Timothy Creazzo, Allen Barnett,
  Christiana Honsberg, and Dennis~W. Prather.
\newblock Thin film silicon solar cell design based on photonic crystal and
  diffractive grating structures.
\newblock {\em Opt. Express}, 16:15238, 2008.

\bibitem{AP85EDP}
Edward~D. Palik, editor.
\newblock {\em Handbook of Optical Constants of Solids}.
\newblock Academic Press, Boston, 1985.

\bibitem{B99SA}
Sadao Adachi, editor.
\newblock {\em Optical Constants of Crystalline and Amorphous Semiconductors}.
\newblock Kluwer Academic Publishers, New York, 1999.

\bibitem{APL06JPP}
J.~P. Prineas, W.~J. Johnston, M.~Yildirim, J.~Zhao, and Arthur~L. Smirl.
\newblock Tunable slow light in bragg-spaced quantum wells.
\newblock {\em Appl. Phys. Lett.}, 89:241106, 2006.

\bibitem{PRB12AF}
A.~Ferreira, N.~M.~R. Peres, R.~M. Ribeiro, and T.~Stauber.
\newblock Graphene-based photodetector with two cavities.
\newblock {\em Phys. Rev. B}, 85:115438, 2012.

\bibitem{OL12MAV}
M.~A. Vincenti, D.~de~Ceglia, M.~Grande, A.~D'Orazio, and M.~Scalora.
\newblock Nonlinear control of absorption in one-dimensional photonic crystal
  with graphene-based defect.
\newblock {\em Opt. Lett.}, 38:3550--3553, 2013.

\end{thebibliography}
\end{document}